\documentclass[aps,prl,preprint]{revtex4}
\usepackage{color}
\usepackage{amsmath}
\usepackage{graphicx} 
\usepackage{bm}
\newcommand{\SB}{\pmb \Sigma}
\newcommand{\OB}{\pmb \Omega}
\newcommand{\xib}{\pmb \xi}
\newcommand{\chib}{\pmb \chi}
\newcommand{\rb}{{\bf R}}
\newcommand{\yb}{{\bf Y}}
\newcommand{\xb}{{\bf X}}

\definecolor{Red}{rgb}{1,0,0} \definecolor{Blu}{rgb}{0,0,1}
\begin{document}
\title{Statistical thermodynamic
  basis in drug-receptor interactions: double annihilation and double
  decoupling alchemical theories, revisited}
\author{Piero Procacci}
\affiliation{Department of Chemistry, University of Florence,
  Italy} \email{procacci@unifi.it} 
\date{today}
\begin{abstract}
Alchemical theory is emerging as a promising tool in the context of
molecular dynamics simulations for drug discovery projects.  In this
theoretical contribution, I revisit the statistical mechanics
foundation of non covalent interactions in drug-receptor systems,
providing a unifying treatment that encompasses the most important
variants in the alchemical approaches, from the seminal Double
Annihilation Method by Jorgensen and Ravimohan [W.L. Jorgensen and
  C. Ravimohan, J. Chem. Phys. 83,3050, 1985], to the Gilson's Double
Decoupling Method [M. K. Gilson and J. A. Given and B. L. Bush and
  J. A. McCammon, Biophys. J. 72, 1047 1997] and the Deng and Roux
alchemical theory [Y. Deng and B. Roux, J. Chem. Theory Comput., 2,
  1255 2006]. Connections and differences between the various
alchemical approaches are highlighted and discussed, and finally placed into the
broader context of nonequilibrium thermodynamics.
\end{abstract}
\maketitle
\section{Introduction}
The determination of the binding free energy in ligand-receptor
systems is the cornerstone of drug discovery.  In the last decades,
traditional molecular docking techniques in computer assisted drug
design have been modified, integrated or superseded using
methodologies relying on a more realistic description of the
drug-receptor system. It has becoming increasing clear that, in order
to reliably rank the affinity of putative ligands for given target, a
microscopic description of the solvent is a crucial ingredient.  As
recently pointed out by Gilson and co-workers,\cite{Nguyen2016} the
nature of the entropic term in binding is intimately related to
microsolvation phenomena in ligand-receptor association that can bring
along very large entropy fluctuations.  

\begin{figure}[h!]
\includegraphics[scale=0.350]{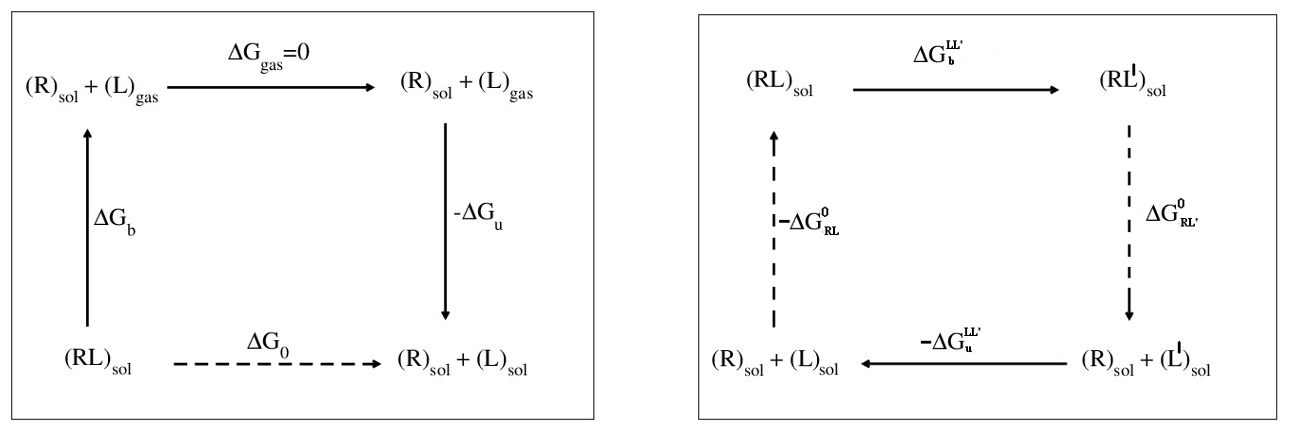}
\caption{The alchemical thermodynamic cycle for computing the absolute
  and relative dissociation free energy, $\Delta G_0$, in
  drug-receptor systems. The subscript ``sol'' and ``gas'' indicates
  solvated and gas-phase species, respectively.  For the alchemical
  determination of absolute {\it standard} free energies (left cycle)
  the ligand must decoupled in the solvated complex and in bulk
  solvent obtaining $\Delta G_0=\Delta G_b -\Delta G_u $.  For
  alchemical determination of {\it relative} standard free energies
  (right cycle) the ligand L must transmuted into the ligand
  L$^\prime$ in the solvated complex and in bulk solvent obtaining
  $\Delta\Delta G_0=\Delta G_b^{\rm LL^\prime} -\Delta G_u^{\rm
    LL^\prime} $.}
\label{fig:cycle}
\end{figure}

In the framework of atomistic molecular dynamics (MD) simulations with
explicit solvent, several computational methods have been devised for
rigorously determining the absolute binding free energy in drug
receptor systems. Most of these methodologies are based on the
so-called alchemical route (see Refs. \cite{chodera11,Roux13} for
recent reviews). In this approach, proposed for the first time by
Jorgensen and Ravimohan\cite{jorgensen85}, the binding free energy is
obtained by setting up a thermodynamic cycle as indicated in Figure
\ref{fig:cycle} and by computing the decoupling free energy of the
ligand in the bound state and in bulk water, indicated hereinafter with
$\Delta G_b$ and $\Delta G_u$, respectively.
These decoupling free energies corresponds to the two closing branches
of the cycle and are obtained by discretizing the alchemical path
connecting the fully interacting and fully decoupled ligand in a
number of intermediate nonphysical states, running for each of these
states equilibrium, fully atomistic molecular dynamics simulations.
Alchemical states are hence defined by a $\lambda$ coupling parameter
entering in the Hamiltonian, varying between 1 and 0 so that at
$\lambda=1$ and at $\lambda=0$ one has the fully interacting and
gas-phase ligand, respectively.  $\Delta G_b$ and $\Delta G_u$ are
usually recovered as a sum of the contributions from each of coupling
parameter windows by applying the free energy perturbation method
(FEP).\cite{Zwanzig54} Alternatively, and equivalently, one can
compute the canonical average of the derivative of the Hamiltonian at
the discrete $\lambda$ points, obtaining the decoupling free energy
via numerical thermodynamic integration (TI).\cite{Kirkwood35}
Finally, the cycle is closed by computing the difference between the
two decoupling free energy along the alchemical path, $\Delta G_b$ and
$\Delta G_u$, obtaining the dissociation free energy in solution.

Gilson et al. \cite{gilson97} criticized Jorgensen's theory by
pointing out that the resulting binding free energies do not depend
upon the choice of standard concentration. In order to define a
reference chemical potential for the decoupling ligand when bound to
the receptor, Gilson introduced a ``restraint'' that somehow keeps the
ligand in the binding place.  This restraint is shown to
yield\cite{gilson97} an additive standard state dependent correction
to the dissociation free energy of $k_BT \ln (V_r/V_0)$, interpreted
as a chemical potential difference of the ligand at concentration
$1/V_r$ and $1/V_0$. According to Gilson, the effect of progressively
strengthening the restraint, leading to a more negative correction,
should be balanced by a larger work integral so that ``errors will
occur only when the integration region defined by the restraint volume
becomes so small that conformations that ought to make important
contributions to the work integral are missed.''  Later Karplus and
co-workers\cite{karplus03} noted that in the final ($\lambda \simeq
0$) stages of the decoupling of the complex, the unrestrained ligand
in DAM may freely rotate and wander to any point in the simulation
system, so that, in order to compute $\Delta G_b$ correctly, the
ligand would have to sample every possible position in the simulation
box, with a standard state correction for the DAM dissociation free
energy equal to the additive term $k_BT \ln (V_{\rm box}/V_0)$ where
$V_{\rm box}$ is the volume of simulation box. In the framework of the
Gilson's DDM theory, these authors hence proposed to enforce a set of
harmonic restraints (with force constant varying from 5 to 50 kcal
mol$^{-1}$[\AA$^{-2}$/rad$^{-2}$]) that restrict both the position and
the orientation of the ligand.  Subsequently, Deng and
Roux\cite{Roux09} proposed a DDM variant whereby the restraints for
the bound state are not present at the end states $\lambda=1$ and
$\lambda=0$ of the alchemical process; rather, they are progressively
switched on and off during the alchemical transformation with a
cancellation effect.  In the Deng and Roux variant, in the limit of
strong restraints, the standard state correction is no longer
dependent on the imposed restraint volume $V_r$. However, it does
requires the estimate of the unknown translational, rotational and
conformational binding site ``volume'' $V_{\rm site}$ in the
complex\cite{Luo02,Deng2006} via an independent unrestrained simulation
of the bound state.

In a series of recent papers, Fujitani and
coworkers,\cite{Yamashita09,Yamashita2014,Yamashita2015} successfully
applied the unrestricted DAM approach to several drug-receptor
systems, in many cases predicting the dissociation free energy via
FEP in close agreement with the experimental values with an average
error of 2/3 kcal mol$^{-1}$. Errors were assessed by repeating
several times the FEP calculations with runs on the order of few ns on
each alchemical states. Most importantly, these authors directly
compared their DAM/FEP values, $\Delta G_{\rm DAM} = \Delta G_b -
\Delta G_u$, to the experimental value $\Delta G_0$, openly criticizing
the DDM standard state correction: ``as far as we know there is no
theoretical or experimental proof that [the standard state corrected]
$\Delta G_0$ meets the definition of the absolute binding energy.[..]
Therefore, we directly compare $\Delta G_{\rm bind}$ with $\Delta
G_0$.''  

The standard state correction issue can be bypassed
altogether by computing relative binding free energies,\cite{Wang2015}
due to the transmutation of a ligand into another in the {\it same
  binding site} and in the solvent. Relative binding free energy
calculations involves as much computations as absolute free energies
do (see Figure \ref{fig:cycle}), and completely neglect the
possibility of a change of binding site volume due to the
transmutation. This approach is hence limited to the assessment of the
binding affinities in strictly congeneric series of ligands with the
tacit assumption of a constant binding site volume upon
transmutation and cannot provide, by any means, a complete tool in
MD-based drug design.

In conclusion, the question of the standard state correction, or,
equivalently, the issue of the binding site volume in drug-receptor
dissociation free energy calculations is either ignored, as in
relative free energy calculations, or treated using methodologies
relying on the definition of arbitrary set of constraints whose
effects on the resulting free energy has never been convincingly
assessed. In any case, the standard state issue, that is indeed
crucial for a reliable MD-based in silico tool in drug discovery, is
still far from being settled.  In this theoretical contribution we
revisit the DAM and DDM theory with a spotlight on the binding site
volume issue, providing a unifying treatment encompassing
Jorgensen, Gilson and Boresch and Roux theories, and finally placing the
alchemical methodology into the broader context of nonequilibrium
thermodynamics.

\section{Achemical theory of non covalent bonding}
Molecular recognignition in host-guest or drug-receptor non covalent
interactions are based on a highly specific molecular
complementary\cite{Lehn1995}, translating in the existence of a
single overwhelmingly prevalent binding ``pose'' defined using an
appropriate set of coordinates that are functions of the ligand and
receptor Cartesian coordinates $x$. A natural coordinate in
ligand-protein binding is represented by the distance ${\bf R}$ of
center of mass (COM) of the ligand with respect to a fixed reference
system with the origin at COM of the protein and oriented along the
inertia axis of the protein.  The vector ${\bf R}$ (in polar
coordinates $r,\theta,\phi$) defines the precise location of the
ligand COM on the protein surface in the bound state. Euler angles can
be further introduced to specify the orientation of the ligand frame
relative to the protein frame. For non rigid ligands and/or binding
pockets, however, a rigorous separation of vibrational and rotational
coordinates is not possible as the inertia tensor of the ligand and,
to a less extent, that of the protein may change significantly upon
binding by coupling to ligand and/or receptor conformational
coordinates. The most general definition of a binding pose is hence
enforced by supplementing the natural coordinate $\rb=r,\theta, \phi$
with an appropriate set of $\xib$ ro-vibrational coordinates defined
with respect to a protein frame in terms of the ligand and receptor
Cartesian coordinates $x$. The set $\{\rb,\xib \}$ should include all
those coordinates whose probability density differ significantly in
going from the bound to the unbound states.

\subsection{Double Decoupling method (DDM)}
In DDM, a set of harmonic restraints are introduced on the ${\bf Y} =
\{ \rb,\xib \}$ $d$-dimensional set of coordinates in order to the keep
the ligand in the binding pose while the decoupling process
proceeds. The easiest way to do so is that of introducing
harmonic potentials for each of these $d$ coordinates, leading to the
restraint potential of the kind
\begin{equation}
V_r(x) =  \frac{1}{2}K_r(r(x)- r_e)^2 + \frac{1}{2}K_\theta(\theta(x)- \theta _e)^2 + 
\frac{1}{2}K_\phi(\phi(x)- \phi _e)^2+ \frac{1}{2}\sum_i^{d-3}  K_\xi^{(i)}(\xi_i(x)-\xi_i^e)^2 
\label{eq:rest}
\end{equation}
The restraint potential can be compactly written in vector notation as
as
\begin{equation}
V_r({\bf Y}-{\bf Y}_e)= \frac{1}{2}({\bf Y} - {\bf Y}_e)^T {\bf K} ({\bf Y} - {\bf Y}_e)
\end{equation}
where ${\bf K}$ is the diagonal matrix of the harmonic force constants. 
Note that the function $e^{-\beta V_r(x)}$ may be interpreted as a
product of independent univariate Gaussian distributions or
equivalently as non normalized multivariate Gaussian distribution in
the $d$ dimensional space defined by the coordinates ${\bf Y}=\{ \rb(x),\xib(x)
\}$
\begin{equation}
e^{-\beta V_r(x)} =  e^{-\frac{1}{2}({\bf Y} - {\bf Y}_e)^T \SB_r^{-1}({\bf Y} - {\bf Y}_e)}   
\label{eq:multi}
\end{equation}
where  the diagonal covariance matrix $\SB_r$ is defined as
\begin{equation}
\SB_r = k_BT  {\bf K}^{-1}
\label{eq:cov}
\end{equation}

As we may not know precisely the geometry of the pose of the ligand in
the binding site, the chosen restraint equilibrium parameters, ${\bf
  Y}_e= \{ \rb_e,\xib_e\}$, can be different from their corresponding true mean values
${\bf Y}_c=\{ \rb_c, \xib_c \}$.  In Ref. \cite{Marsili10}, in the context of
single molecule pulling experiments, a simple relation was derived
between the free energy of the driven system (i.e. with Hamiltonian
including the harmonic potential of an external device coupled to a
specific molecular distance $R$) and the free energy of the system
with unperturbed Hamiltonian along the driven coordinate (i.e. the
potential of mean force along $R$). The relation proposed by Marsili
(Eq. 7 in Ref. \cite{Marsili10}) can be straightforwardly applied to
any of the restrained $\lambda$ alchemical state in DDM as:
\begin{equation}
G_r(\SB_{r},\yb_e,\lambda) = G(\yb,\lambda)  + V_r({\bf Y}
-{\bf Y}_e) + 
k_{\rm B} T \ln \left ( \frac{P(\yb |\SB_r ,\yb_e,\lambda)}{P(\yb_*)} \right )
\label{eq:reweight}
\end{equation}
where 
\begin{eqnarray}
G_r(\SB_r,\yb_e,\lambda) & = & -k_BT \ln \left [ {\cal C} \int dx e^{-\beta[H(x,\lambda) +
    V_r ({\bf Y}  -{\bf Y}_e)]} \right ] \label{eq:Z}\\
 G(\yb,\lambda) & = & -k_BT \ln \left [ \frac{\int dx \delta (\yb-\yb(x))
  e^{-\beta H(x,\lambda)}}{\int dx \delta (\yb_* -\yb(x))
  e^{-\beta H(x,\lambda)}} \right ] = -k_bT \ln
 \frac{P(\yb)}{P(\yb_*)}
\label{eq:Zm}
\end{eqnarray} 
Here, $G_r(\SB_r,\yb_e,\lambda)$ is the free energy of the
restrained system (${\cal C}$ is an $h$ dependent constant that
makes argument of the logarithm adimensional) and
$G(\yb,\lambda)$ is the free energy of the unrestrained system at
${\bf Y}=\rb,\xib$ with respect to some immaterial reference state
at ${\bf Y}_*=\rb_*,\xib_*$  In Eqs \ref{eq:Z}
and \ref{eq:Zm}, $H(x,\lambda)$ is the Hamiltonian at the alchemical
state $\lambda$, with $x$ encompassing all solvent, ligand and
receptor coordinates. $P(\yb |\SB_r, \yb_e,\lambda) \equiv \langle
\delta(\rb-\rb(x))\delta (\xib -\xib(x))\rangle_r $, finally, is the
canonical probability density evaluated at ${\bf Y}=\{ \rb,\xib \}$
for the restrained system with free energy given by Eq. \ref{eq:Z}.

In the
alchemical simulation of the complex, one computes, either via FEP or
TI, the free energy difference between the states at $\lambda=1$
(interacting ligand) and $\lambda=0$ (gas-phase ligand), subject to the
restraint potential $V_r$, Eq. \ref{eq:rest}.  In force of Eq. \ref{eq:reweight}, we
therefore get the $\yb_*$ independent relation  
\begin{eqnarray}
\Delta G_r(\SB_r,\rb_e,\xib_e) & = & G_r(\SB_r,\rb_e,\xib_e,0)-G_r(\SB_r,\rb_e,\xib_e,1)  \nonumber \\
& = &  
\Delta G(\rb,\xib)
+ k_{\rm B} T \ln \frac{P(\rb,\xib |\SB_r, \rb_e, \xi_e,0)} {P(\rb,\xib |\SB_r, \rb_e, \xib_e,1)}
\label{eq:reweight2}
\end{eqnarray}
where I have used the expanded notation for ${\bf Y}=\{ \rb,\xib \}$
and where 
\begin{equation}
\Delta G(\rb,\xib) =-k_BT \ln \left [ \frac{\int dx \delta (\rb-\rb(x))\delta (\xib-\xib(x))
  e^{-\beta H(x,0)}}{\int dx \delta (\rb-\rb(x))\delta (\xib-\xib(x))
  e^{-\beta H(x,1)}} \right ]
\label{eq:decou}
\end{equation}
is the decoupling free energy of the unrestrained system evaluated at
${\bf Y}=\{ \rb,\xib \}$ and where $\Delta G_r(\SB_r,\rb_e,\xi_e)$ corresponds to
decoupling free energy of the restrained complex.  Note
that, since there is no change in the parameters ${\bf Y}_e= \{\rb_e,\xi_e \}$
in going from the initial (coupled) to the final (decoupled) state,
there can't be correspondingly no change in the harmonic potential
energy at $\{\rb,\xib\}$ due to the restraint.

The $\xib$-dependent {\it decoupling free energy of the unbound state}
can be defined as\cite{Roux09}
\begin{eqnarray}
\Delta G_u(\xib)
& = &
- k_BT \ln \left [ \frac 
{\int dx \delta(\rb_{\infty}-\rb(x))\delta (\xib-\xib(x)) e^{-\beta H(x,0)}} 
{\int dx \delta(\rb_{\infty}-\rb(x))\delta (\xib-\xib(x)) e^{-\beta H(x,1)}} 
\right ] 
\label{eq:dgu}
\end{eqnarray}
where, $\rb_{\infty}$ represents a ligand-receptor COM distance that
is large enough to allow the ligand and the receptor to interact only
with the solvent when $\lambda \ne 0$. $\Delta G(\xib)$ represents the
reversible work to bring the unbound ligand {\it and} unbound receptor
(set at a relative vector distance $\rb_{\infty}$ and in the ro-vibrational
states defined by the vector $\xib$) from the bulk into the gas-phase. This work may depend
on the $\xib$ coordinates in case of, e.g., competing conformational
states of the ligand and/or protein involved in the binding.  For a
rigid ligand and rigid binding pose, $\xib$ can be taken to coincide
with the three Euler angles, $\OB$, defining the orientation of the
ligand with respect to the protein frame.  In this case, all
rotational states at $\rb_{\infty}$ (i.e. for the unbound or free
ligand) have equal weights so that $\Delta G_u(\OB) \equiv \Delta G_u$
is independent of $\OB$.  In DDM theories, while ligand conformational
changes upon binding may\cite{Roux04} or may not\cite{karplus03}
accounted for, the fact that the receptor may change as well its
conformational state in the binding process is generally overlooked.
By subtracting Eq. \ref{eq:dgu} in Eq. \ref{eq:decou}, we obtain
\begin{eqnarray}
\Delta G_r(\rb,\xi) - \Delta G_u (\xib) & = & 
-k_BT \ln \left [ \frac{\int dx \delta (\rb-\rb(x))\delta (\xib-\xib(x))
  e^{-\beta H(x,0)}}{\int dx \delta (\rb-\rb(x))\delta (\xib-\xib(x))
  e^{-\beta H(x,1)}} \right ] +   \nonumber  \\
& + & 
k_BT \ln \left [ \frac 
{\int dx \delta(\rb_{\infty}-\rb(x))\delta (\xib-\xib(x)) e^{-\beta H(x,0)}} 
{\int dx \delta(\rb_{\infty}-\rb(x))\delta (\xib-\xib(x)) e^{-\beta H(x,1)}} 
\right ]
\nonumber \\
& = &  
-k_BT \ln \left [ \frac{\int dx \delta (\rb_{\infty}-\rb(x))\delta (\xib-\xib(x))
  e^{-\beta H(x,1)}}{\int dx \delta (\rb-\rb(x))\delta (\xib-\xib(x))
  e^{-\beta H(x,1)}} \right ]  \nonumber \\
& = & 
 -w(\rb,\xib)  
\label{eq:pmf}
\end{eqnarray}
where we have exploited the fact that the probability densities of the
decoupled ligand and receptor ($\lambda=0$) with respect to ${\bf R}$
is uniform.  $w(\rb,\xib)$ on the rhs of Eq. \ref{eq:pmf} represents
the reversible work, or potential of mean force, for bringing a
separated ligand and receptor in the $\xib$ ro-vibrational arrangement
into the corresponding bound conformation at $\rb$.

If in Eq. \ref{eq:reweight2} and Eq. \ref{eq:pmf} we choose
$\xib_e=\xib_c$ and $\xib=\xib_c$ and we use Eq. \ref{eq:pmf}, we
obtain
\begin{equation}
\Delta G_r(\SB_{r},\rb_c,\xib_c)
- \Delta G_u(\xib_c) = -w(\rb_c,\xib_c)  + 
k_{\rm B} T \ln \frac
{
P(\rb,\xib |\SB_r, \rb_c, \xib_c,0)
}
{
P(\rb,\xib |\SB_r, \rb_c, \xib_c,1)
}
\label{eq:reweight3}
\end{equation}
Eq. \ref{eq:reweight3} expresses the fact that the 
dissociation free energy with a set of harmonic restraints of the kind
of Eq.\ref{eq:rest} computed in DDM simulation via FEP or TI, namely the quantity 
\begin{equation}
\Delta G_d(\SB_{r},\rb_c,\xib_c) = \Delta G_r(\SB_{r},\rb_c,\xib_c)
- \Delta G_u(\xib_c)
\label{eq:DGD} 
\end{equation}
is equal to minus the drug-receptor PMF at $\rb_c,\xib_c$ plus a
correction related to the logarithm of the ratio of the canonical
probability distributions for the {\it restrained} decoupled and
coupled bound states, respectively, evaluated in both cases at the
same point $\rb_c , \xib_c$.  I stress that for
Eq. \ref{eq:reweight3} to be valid, the canonical probabilities at the
end states, $P(\rb,\xib |\SB_c, \rb_c, \xib_c,1)$ and $P(\rb,\xib
|\SB_c, \rb_c, \xib_c,0)$, must be both evaluated with the restraint
in place.  

\begin{figure}[h!]
\includegraphics[scale=0.5]{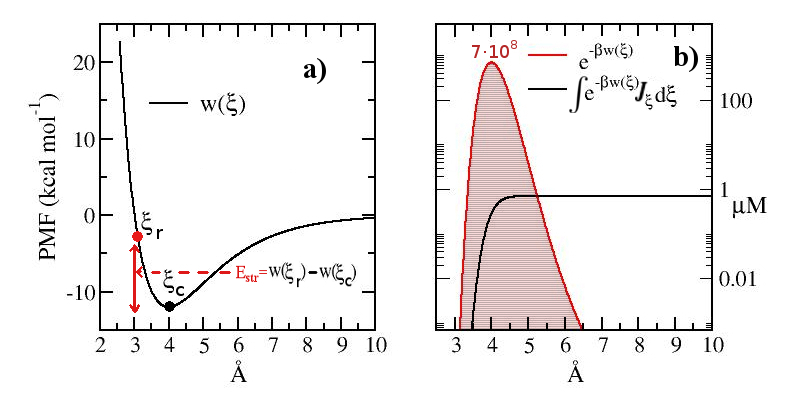}
\includegraphics[scale=0.5]{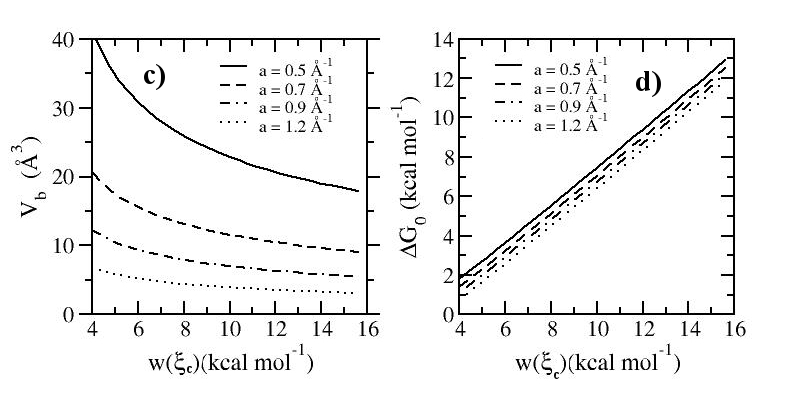}
\caption{Relation between $V_b$ and the PMF $w(\xi)$ for a simple
  monoatomic ligand ($\xi=R,\theta,\phi$) {\bf a}: PMF as a function
  of the ligand-receptor distance. The PMF is modeled with a distance
  dependent Morse potential of the form $w(r)= D[1- e^{-a(r-r_0)}]^2
  -D$.  The $\theta,\phi$ dependency is such that bonding may occur
  only in a solid angle of $D_{\Omega}=\pi/2$ corresponding to one
  octant of the $4\pi$ integrated orientational space. The red segment
represents the strain energy due to a wrong choice of the restraints 
(see text). {\bf b}:
  adimensional factor $e^{-\beta w(\xi)}$ as a function of the
  ligand-receptor distance (maximum value at $\xi=\xi_c$). The shaded
  area defines the volume $V_b$. The integral of the function
  $e^{-\beta w(\xi)} J_{\xi}$ ($J_\xi$ is the Jacobian of the
  transformation $\xi=\xi(x)$ within the shaded area is the solid
  black line and yields the equilibrium constant $K_{\rm eq}$
  (reported in $1/\mu$ M units). $V_b$ ({\bf c}) and standard
  dissociation energy ({\bf d}) as a function of $w(\xi_c)$ for
  various $a$ values (width) of the Morse potential.}
\label{fig:pmf}
\end{figure}

How does then the FEP or TI computed DDM dissociation free energy
$\Delta G_d(\SB_{r},\rb_c,\xib_c)$ relate to the {\it standard}
dissociation free energy $\Delta G_{d0}$? Or, equivalently, how does
the potential of mean force $w(\rb_c,\xib_c)$ at its minimum value
$\{\rb, \xib \}=\{\rb_c,\xib_c\}$ relate to the dissociation constant
$K_d/C_0=e^{-\beta\Delta G_{d0}}$?  I recall that in the present
treatment, the $\{\rb, \xib \}$ coordinates are defined with respect
to the fixed inertia system of the receptor. It is convenient to
further distinguish between rotational coordinates of the ligand
relative to the receptor and all other (ligand and receptor)
conformational coordinates involved in the definition of the complex,
namely $\xib\equiv \OB,\chib$.  While the rotational states $\OB$
defining the orientation of the ligand frame relative to the fixed
protein have all equal probability of $1/8\pi^2$ when the molecules
are separated in the bulk (no matter what the conformational states of
the partners are), the $\chib$ conformational coordinates of the
separated species in standard conditions can be rationalized in terms
of {\it conformational} basins with uneven weights.  It can then be
shown that the dissociation constant in the {\it infinite dilution limit}
for a fixed conformation $\chib$ is given by\cite{Luo02}
\begin{eqnarray}
\frac{1}{K_d(\chib)} & = & \frac{1}{8\pi^2}
\int_{{\cal D}_b(\chib)} e^{-\beta w(\rb,\OB,\chib)} d\rb d \OB \\
& = & \frac{V_b(\chib)}{8\pi^2} e^{-\beta w(\rb_c,\OB_c,\chib)}
\label{eq:vsite}
\end{eqnarray}
where the integration domain, ${\cal D}_b(\chib)$, must be restricted to
the region of existence of the complex between the receptor and the
ligand in the fixed conformational states defined by the $\chib$
coordinates.\cite{gilson97,Luo02} In the second equality we have
written the integral (that has the dimension of a volume and square
radiants) in terms of an effective volume $V_b(\chib)$ times the
potential of mean force at the bottom of the well, $w({\bf
  R}_c,\OB_c,\chib)$.  The physical meaning of such volume is
schematically illustrated in Figure \ref{fig:pmf} for a simple
monoatomic ligand.
Here, we have assumed a single minimum PMF of the kind $w({\bf
  R})=D_e(r)\Omega(r,\theta,\phi)$, where $D_e(r)$ is a Morse
potential and $\Omega(r,\theta,\phi)$ is an appropriate square well
potential defining the entrance angle of the monoatomic ligand into
the binding pocket. Note that (Figure \ref{fig:pmf}{\bf b}) the integral defining the equilibrium
constant can be extended beyond the ${\cal D}_b$ domain with no
appreciable change in $K_{\rm eq}$. 

Returning back to the general Eq. \ref{eq:vsite}, for a polyatomic
ligand, $V_b(\chib)$ also
includes a rotational contribution due to the librations of the ligand
in the pocket, \cite{gilson97} 
when the ligand and the receptor are in the given conformational state $\chib$.  
We can approximate the integrand in
Eq. \ref{eq:vsite} with respect to the coordinates $\xb = \rb,\OB$
with a multivariate Gaussian distribution of appropriate covariance
$\SB_r$ (see Figure \ref{eq:pmf} (b)), i.e.
\begin{equation}
\int_{{\cal D}_b(\chib)} e^{-\beta w(\rb,\OB,\chib)} d\rb d \OB = 
e^{-\beta w(\rb_c,\OB_c,\chib)} \int  e^{-\frac{1}{2}({\bf X} - {\bf X}_c)^T \SB_b^{-1}(\chib)({\bf X} - {\bf
    X}_c)} d\xb   
\label{eq:multi}
\end{equation}
so that 
\begin{equation} 
V_b(\chib) = \sqrt{2\pi^3 |\SB_b(\chib) | } 
\label{eq:vb}
\end{equation}
Going back to Eq. \ref{eq:vsite}, the overall
dissociation constant can be calculated as a standard canonical
average:
\begin{equation}
\frac{1}{K_d} = \int d\chib P(\chib)  \frac{1}{K_d(\chib)} = \int d\chib P(\chib)\frac{V_b(\chib)}{8\pi^2} e^{-\beta w(\rb_c,\OB_c,\chib)} 
\end{equation}
In the infinite dilution limit, the probability density $P(\chib)
\equiv \langle \delta(\chib -\chib(x) \rangle$ is identical to the
probability density of the conformational coordinates $\chib$ of the
{\it separated} species. If the binding involves only one
conformational state or basin defined by the integration domain 
${\cal  D}_{\chi}$ of the ligand and the receptor, i.e. if
$K_{d}(\chib)$ is overwhelmingly dominated by $\chib \in D_{\chi}$,  we obtain
\begin{equation}
\frac{1}{K_d} = 
\int_{{\cal D}_\chi}  d\chib P(\chib)  \frac{1}{K_d(\chib)} =
W(\chib_c) \frac{V_b(\chib_c)}{8\pi^2} e^{-\beta w(\rb_c,\OB_c,\chib_c)}
\label{eq:kd}
\end{equation}
where we have defined the mean (adimensional) conformational weight
$W(\chib_c)$.  If the conformational states spanned by the $\chib$
coordinates are well separated and characterized by deep minima, then
$W(\chib_c)$ can be identified, in first instance, with the canonical
weight in dilute solution of the binding ligand/receptor conformation
{\it for the separated species}. If such binding conformation has a
low weight for the separated species, then it means that the drug
and/or the receptor experiences substantial conformational changes
upon binding and that the free energy gain in the association  process
comes either from the volume (or entropy, {\it vide infra})  term $V_b(\chib)$ or from the
enthalpic gain due to the $e^{-\beta w(\rb_c,\OB_c,\chib_c)}$ term. 

Taking into account that $\Delta G_{d0}=-k_BT
\ln(K_dV_0)$ Eq. \ref{eq:kd} can be equivalently written in terms of
dissociation free energy as
\begin{equation}
\Delta G_{d0} =  -w({\bf R}_c,\OB_c ,\chib_c) + 
k_BT \ln \left ( \frac{V_b(\chib)}{8\pi^2 V_0} \right ) + k_BT \ln W(\chib_c) 
\label{eq:dg0}
\end{equation} 
Again, note that while the vector distance ${\bf R}$ is a collective
variable (CV) bearing no coupling with other ligand-receptor CVs, the
integration domain of the $\OB$ CV in the bound state is in principle
dependent on the conformational state $\chib$. 
In DDM theory, it is
tacitly assumed that the $\chib$ conformational coordinates pertain
the ligand only (i.e. the conformational state of the receptor is
invariant upon binding) and that the orientational volume spanned by
the ligand relative to the receptor in the binding site is
approximately independent of the conformational state of the system.
In this rather strong assumption, that can be in essence identified
with the rigid rotor harmonic oscillator (RRHO) approximation, the
determinant in Eq. \ref{eq:vb} is diagonal and the  
volume $V_b(\chib)$ can be written as product of a $\chib$ independent
translational volume $V_{t}$ and an orientational volume $V_{\Omega}$
(expressed in radiants) leading to the expression
\begin{equation}
\Delta G_{d0}^{[{\rm RRHO]}} =  -w({\bf R}_c,\OB_c ,\chib_c) + 
k_BT \ln \left ( \frac{V_t}{V_0} \right ) + 
k_BT \ln \left ( \frac{V_\Omega}{8\pi^2} \right ) + 
k_BT \ln W(\chib_c)
\label{eq:rrho} 
\end{equation} 
One can see the three logarithmic terms in Eq. \ref{eq:rrho} as a
translational, rotational and conformational entropy loss of the bound
state, producing a penalty in the binding affinity, thus writing
Eq. \ref{eq:rrho} in the familiar form
\begin{equation}
\Delta G_{d0} = \Delta H_d -T\Delta S_{d0} 
\end{equation}
with the dissociation enthalpy $\Delta H_d=-w({\bf
  R}_c,\OB_c,\chib_c)$ given by the PMF at the bottom of the {\it
  single} well in the ${\bf R}, \OB, \chib $ space and the standard
state dependent and volume related dissociation entropy $\Delta
S_{d0}= -k_B \left [ \ln \left ( \frac{V_t}{V_0} \right ) + \ln \left
  ( \frac{V_{\Omega}}{8\pi^2} \right ) + \ln W(\chib_c) \right ]$.
Hence, the more tightly is bound the ligand in the pocket, the smaller
will be the ``volumes'' $V_t$, $V_{\Omega}$ and $W(\chib)$ and the
larger is the entropy loss due to association.   

Incidentally, we may hence say that Eq. \ref{eq:rrho} constitutes the
statistical mechanics foundation of the Docking approach, essentially
based on the underlying RRHO approximation.  If, for example, we
assume that $N_c$ represents a set of equally populated conformational
states of the free ligand (due to, e.g., rotable
bonds\cite{Chang2007}), Eq. \ref{eq:rrho} may be rearranged
\begin{equation}
\Delta G_{d0}^{[{\rm RRHO]}} =  -w({\bf R}_c,\OB_c ,\chib_c) + 
k_BT \ln \left ( \frac{V_t}{V_0} \right ) + 
k_BT \ln \left ( \frac{V_\Omega}{8\pi^2} \right ) 
- k_BT \ln N_c
\label{eq:rrho2} 
\end{equation} 
In molecular Docking, the energetic contribution,
$\Delta H = -w({\bf R}_c,\OB_c ,\chib_c)$, is evaluated using
molecular mechanics Poisson-Boltzmann surface area
(MM/PBSA))\cite{Case1998,Kollman2000} or the molecular
mechanics generalized Born surface area
(MM/GBSA)\cite{Kollman2000,Greenidge2013,agbnp} models, while the
  elusive volume entropic contributions, $k_BT ( \ln \left (
  \frac{V_t}{V_0} \right ) + \ln \left ( \frac{V_\Omega }{8\pi^2}
  \right )$, are either evaluated using MD
  methodologies\cite{Hou2011} or by simplified analytical
  estimates.\cite{Procacci2016}

Going back to Eq. \ref{eq:reweight3}, Eq. \ref{eq:dg0} provides the
searched relationship between the potential of mean force $ w({\bf
  R}_c,\xib_c)= w({\bf R}_c,\OB_c ,\chib_c)$ and the standard
dissociation free energy $\Delta G_{d0}$ in the context of DDM theory.
If we use Eq. \ref{eq:dg0} in Eq. \ref{eq:reweight3} and using the
definition Eq. \ref{eq:DGD}, we finally find
\begin{equation}
\Delta G_{d0} = \Delta G_d(\SB_{r},\rb_c,\xib_c) + k_BT \ln \left ( 
\frac{V_b(\xib_c)}{8\pi^2 V_0} \right ) -  
k_{\rm B} T \ln \frac
{
P(\rb,\xib |\SB_r, \rb_c, \xib_c,0)
}
{
P(\rb,\xib |\SB_r, \rb_c, \xib_c,1)
}
\label{eq:reweight4}
\end{equation}
where we have re-defined the overall binding site volume as  
\begin{equation}
V_b(\xib_c) = V_t V_\Omega(\chib_c)    W(\chib_c)
\label{eq:vb2}
\end{equation} 

Equation \ref{eq:reweight4} defines a DDM {\it general} relation
embracing (as we shall see further on) all current binding theories
from the DAM approach with no restraints to the Deng and Roux method
with strong restraints.  Note again that, in the general case, the
``rotational volume'', $V_\Omega(\chib_c)$, is a function of the
conformational states.
\subsubsection{Boresch's and Deng's theory: stiff restraint regime}
When $K_i \rightarrow \infty$, i.e. in the so-called stiff-spring
regime\cite{Park04,Marsili10}, the last logarithmic term on the rhs of
Eq. \ref{eq:reweight4} is zero since the probability densities for the
restrained system in the $\lambda=1$ and $\lambda=0$ states becomes
identical. According to eq. \ref{eq:reweight3}, the alchemically
determined dissociation free energy (Eq. \ref{eq:DGD}), $\Delta
G_d(\SB_{r},\rb_c,\xib_c)$, can be thus taken to be equal to minus the
PMF at $\{ {\bf R}, \xib \} =\{ {\bf R_c},\xib_c \}$, i.e.
\begin{equation}
\Delta G_d(\SB_{r},\rb_c,\xib_c) =  - w ( {\bf R_c},\xib_c )  
\label{eq:strong}
\end{equation}
Consequently, in order to recover the dissociation standard free
energy in alchemical simulations with strong restraints, the strong
restraint $ \Delta G_d(\SB_{r},\rb_c,\xib_c)$ free energy should be
corrected by a volume term $V_b(\xib_c)$ that, in the limit of large
force constants $K_i$, is {\it independent} on $\SB_r$ and is
related to the unknown binding site volume $V_{\rm site}$, i.e. 
\begin{equation}
\Delta G_{d0} =   
\Delta G_d(\SB_{r},\rb_c,\xib_c)
 + k_BT \ln \left (
\frac{V_b(\xib_c)}{8\pi^2 V_0} \right ) 
\label{eq:dg0a} 
\end{equation}
$V_b(\xib_c) $ can be taken as a system-dependent volume defined by
the domain ${\bf R}, \OB$ ${\cal D}_b(\chib)$ for the bound state when
the ligand and the receptor are in the $\chib$ conformational states.
It important to stress that the size and the units of the volume
$V_b(\xib_c)$ {\it depends on the choice of the ro-vibrational $\xib$
  coordinates used to define the binding site}.  Provided that
$V_b(\xib_c) $ can be somehow estimated in independent {\it
  unrestrained} simulations of the free ligand (needed for measuring
$W(\chib_c)$) and of the complex, Eq. \ref{eq:dg0a} allows to compute
the absolute dissociation free energy from the difference of the
decoupling free energies of the free ligand and of complex obtained by
FEP or TI, where the latter is tightly kept around the $\rb_c,\xib_c$
ligand-receptor position by a set of strong restraints of the form
Eq. \ref{eq:rest}. Eq.  \ref{eq:dg0a} was previously derived using a
different route by Boresch {\it al}\cite{karplus03} and by Deng and
Roux\cite{Deng2006}. In the strong restraint approach, the estimate of
the dissociation free energy crucially depends on the estimate of the
binding site volume $V_b(\xib_c)$ that can vary by several kcal
mol$^{-1}$,\cite{Deng2006} hence spanning more than three orders of
magnitude in the inhibition constant. Moreover, the parameters
$\xib_e$ in the restraint potential, Eq. \ref{eq:rest}, should be chosen such that they coincides with the
corresponding mean values of the unrestrained bound state
$\xib_c=\langle \xib \rangle_b$, where the subscript $b$ indicate that
the mean must be taken over bound state canonical configurations.  If
any of the $\{ \xi_i^e \}$ differs from the corresponding equilibrium
value $\{ \xi_i^c \}$, then, as shown in Figure \ref{fig:pmf}, the
system is subject to a strain potential that will be reflected in the
PMF and hence on $V_b$. Probably, the major weakness in DDM with
strong restraints lies in the choice of the restrained coordinates
themselves, that impact on the size and units of $V(\xib_c)$.  First
of all, the number and the nature of the ligand and receptor
conformational coordinates participating to binding is not known from
the start. Secondly, whatever their choice, due to the inherent
fluxional nature\cite{Gilson2010} of ligands and receptor, these
coordinates will be coupled to other ligand and receptor coordinates
so that restraining them may prevent the sampling of configurational
states that are relevant for the binding affinity. In some sense,
Boresch and Deng theory appears essentially to be based on the
traditional picture of ``lock and key'' model\cite{Fischer1894} for
binding, with a systematic underestimation of the binding site volume
$V_b(\xib_c)$ due to the neglect of any effect of receptor and ligand
conformational reshaping (``induced fit'' model\cite{Koshland1958}).

\subsubsection{Gilson's theory: Intermediate restraint regime} 
We now
assume that we impose only translational and orientational restraints
and that these restraint are weak enough to allow the ligand-receptor
system, to canonically sample all $\chib$ conformational states that
are important for binding. This can be practically achieved, for example, by using
only rigid portions of the ligand and the receptor in order to define
the relative ligand-receptor orientation $\OB$ with a possibly
negligible impact on the sampling of conformational states.  At the
same time the translational restraint potential should be strong
enough to prevent the ligand to freely drift away from the binding
site at any $\lambda$ alchemical states. In this case, we can identify
$\xib_c$ with $\OB_c$ so that we may write  the probability density
of the decoupled restrained bound state as 
\begin{eqnarray}
P(\yb|\SB_r, \yb_c,0) & =  & P(\rb,\OB |\SB_r, \rb_c, \OB_c,0) \nonumber \\ 
   & = & \frac{1}
{\int 
e^{-\frac{1}{2}({\bf Y} - {\bf Y}_c)^T \SB_r^{-1}({\bf Y} - {\bf Y}_c)}   
d\yb} = \frac{1}{V_r}  
\label{eq:vr}
\end{eqnarray} 
where we have used Eqs \ref{eq:rest} and Eq. \ref{eq:cov} and where
$V_r$ defines the temperature dependent allowance restraint volume
such that $V_r > V_b$. The probability
density of the fully coupled restrained system can be written as a
product of two multivariate Gaussian distribution with covariance
matrix $\SB^{-1}=\SB_b^{-1}+\SB_r^{-1}$ defined in the $\{ \rb,\OB \}$ space, i.e.
\begin{eqnarray}
P(\yb|\SB_r, \yb_c,1) & = & \frac{ e^{-\beta w(\yb_c)}}{
  \int
  e^{-\beta [ w(\yb) + V(\yb -\yb_c)]  } d\yb } \nonumber \\
& = &  \frac{1} 
{\int e^{-\frac{1}{2}(\yb-\yb_c)^T  \SB_b^{-1}(\yb-\yb_c)}   e^{-\frac{1}{2}(\yb-\yb_c)^T 
      \SB_r^{-1}(\yb-\yb_c)} d\yb}   
\nonumber   \\
& = & \frac{\sqrt{2\pi^d \det[\SB_r + \SB_b]
}}{V_rV_b} = \frac{\sqrt{\det[{\bf 1} + \SB_r^{-1} \SB_b]}}{V_b} 
\label{eq:veff}
\end{eqnarray} 
where $V_b = \int_{{\cal D}_b} e^{-\beta w(\yb)} d\yb \simeq \int
e^{-\frac{1}{2}(\yb-\yb_c)^T \SB_b^{-1}(\yb-\yb_c)} d\yb$ and where
the effective covariance $\SB_b$ no longer depends on the
conformational states, whose contribution is supposed to be
implicitly integrated away in the PMF $w(\rb,\OB)$. 
Inserting Eqs. \ref{eq:veff} and \ref{eq:vr}
into Eq. \ref{eq:reweight4}, we find
\begin{equation}
\Delta G_{d0} = \Delta G_d(\SB_{r},\rb_c,\OB_c)   + k_BT \ln \left ( 
\frac{V_{\rm r}}{8\pi^2 V_0} \right ) + k_BT \ln  \sqrt{\det({\bf 1} + \SB_r^{-1} \SB_b)}
\label{eq:gilson} 
\end{equation}
In the assumption that the last term is small and can be neglected
(i.e. $V_r \gg V_b$), and factoring the restraint volume $V_r$ in
translational and orientational parts $V_I$, $\xi_I$, then
Eq. \ref{eq:gilson} is identical to the Equation proposed by
Gilson.\cite{gilson97} I stress that Eq. \ref{eq:gilson} was derived
by introducing ligand-protein rotational coordinates that are supposed
to be decoupled from any conformational state, so that
$w(\rb_c,\OB_c)$ represents the reversible work to bring the ligand
form the bulk state to the bound state defined by the coordinates
$\rb_c,\OB_c$, irrespective of the conformational states. 
DDM with weak restraint potentials should be
handled with due care by practitioners. In case of highly symmetric
ligands like benzene in T-lysozime,\cite{Deng2006} for example, 
weak orientational restraints may prevent the sampling of the bound
conformations that are defined by a mere exchange of the atom labels
due to rotational operations of the symmetry group of the ligand (say
$\sigma$), underestimating the {\it conformational volume} in the
bound state and hence the dissociation free energy. If the weak
orientational restraints prevents the sampling of any of the
equivalent $\sigma=12$ states of benzene, then the free energy should
be corrected by an additive term $k_BT\ln 12$ apparently due to
``symmetry''.  If instead the restraints are engineered so that they
allow the sampling of the bound states generated by rotations around
the six-fold axis of the benzene molecule but not of those that can be
generated by rotation around the 2-fold symmetry axis, then the
correction factor reduces to $k_BT \ln 2$. Incidentally, I remark
that this kind of corrections applies only to DDM with weak restraints
and not to the Boresch and Deng variant with strong restraints,
provided that in the $V_b(\chib)$ measure of the binding site volume
for the {\it unrestrained} system all relevant conformational states
have sampled.

As discussed in Ref. \cite{gilson97}, for
Eq. \ref{eq:gilson} to hold, it must be that
\begin{equation}
  \frac{\partial \Delta G_d(\SB_{r},\rb_c,\OB_c) } {\partial V_r} =
  -\frac{k_BT} {V_r } 
\label{eq:deriv}
\end{equation}
where, in taking the derivative, we have neglected the last term in
Eq. \ref{eq:gilson}.  Eq. \ref{eq:deriv} provides in principle a mean
to assess whether the chosen restraints obeys the Gilson's regime.  In
fact, by computing the uncorrected alchemical dissociation free energy
for different restraint potentials at constant temperature and
pressure and plotting the result as a function of $-1/V_r$ we should
find a straight line with slope of $k_BT$.

\subsubsection{Jorgensen's  theory: Unrestrained (DAM) regime.}
 What happens when instead we let
${\bf K} \rightarrow 0$ in Eq. \ref{eq:reweight3}? In this case, as
first remarked in Ref. \cite{karplus03}, the alchemical procedure
becomes cumbersome since the standard dissociation free energy should
be in principle recovered by the single {\it equilibrium} simulation
at the fully coupled state $\lambda=1$. The dissociation free energy
detected in the unrestrained simulation depends on the nominal
concentration of the species imposed by the periodic boundary
conditions (PBC), i.e.  on the MD box volume $V_{\rm box}$.  The
fraction of dissociated species can be expressed as function of the
ratio $r=K_d/C_{\rm box}$, where $C_{\rm box}=1/V_{\rm box}$ is the
nominal concentration imposed by the PBC, as
\begin{equation}
f=\frac{r}{2} \left [ \left (1+\frac{4}{r} \right )^{1/2} -1  \right ]
\label{eq:frac}
\end{equation}
Note that in the high concentration limit we have that $\lim_{r \rightarrow 0} f =
0$ while at infinite dilution $\lim_{r \rightarrow 0} f = 1$.
In simulations of typical drug-receptor systems,  $V_{\rm
  box}$ may be taken to vary in the range 10$^5$:10$^6$ \AA$^3$. Hence,
for a micromolar to nanomolar ligand,  $1/K_d$ 
varies in the range 10$^9$:10$^{12}$ \AA$^{3}$ so that the ratio $r$ is 
of the order of 10$^{-7}$:10$^{-3}$.  In this conditions, we have that
$f=r^{1/2} + o(r^2)$ and the box dependent free energy evaluated in the
equilibrium simulation at $\lambda=1$ may be computed as 
\begin{eqnarray}
\Delta G_d(V_{\rm box}) & = &  -k_BT \ln( K_d V_{\rm box}) = -k_BT\ln \frac{f^2}{1-f}   \nonumber  \\
&= & - k_BT \ln r + k_B T\ln (1-r^{1/2})   \nonumber  \\
& \simeq & \Delta G_{d0} -  k_BT \ln \frac{V_{\rm box}}{V_0}
\label{eq:fraction}
\end{eqnarray}
where in the last equation we have neglected the quantity $k_BT
\ln(1-r^{1/2})\simeq -k_BT r^{1/2}$ and exploited the fact that
$\Delta G_{d0}=-k_BT \ln (K_d V_0)$. The standard free energy can hence be
determined by a single very long simulation at the fully coupled state
using Eq. \ref{eq:fraction}. However, one can also choose to implement
the cumbersome alchemical methodology in the unrestrained
version, by applying the ${\bf K} \rightarrow {\bf 0}$ limit of the
general Equation \ref{eq:reweight3} and assuming that only a restraint
on $\rb$ is imposed, i.e. 
\begin{equation}
\lim_{\bf K \rightarrow 0} \Delta G_r({\beta \bf K}^{-1},\rb_c)
- \Delta G_u = -w(\rb_c)  + k_{\rm B} T  
\ln \left [ \frac
{
\lim_{\bf K \rightarrow 0}  P(\rb,|{\beta \bf K}^{-1}, \rb_c, 0)
}
{
\lim_{\bf K \rightarrow 0}  P(\rb |{\beta \bf K}^{-1}, \rb_c,1)
} \right ] 
\label{eq:COM}
\end{equation}
where 
\begin{equation}
\Delta G_{d0} =  -w({\bf R}_c) +  k_BT \ln \left ( \frac{V_T}{V_0} \right )
\label{eq:dg0com}
\end{equation}
$V_T = \int_{{\cal D}_b} e^{-\beta w(\rb) } d\rb $ is the allowance
oscillation volume of the COM vector distance ${\bf R}$ in the complex
irrespective of the ligand-receptor orientational and conformational
coordinates. In the limit ${\bf K \rightarrow 0}$ , the restraint the
probability density of the decoupled system is given by
\begin{equation}
\lim_{\bf K \rightarrow 0}  P(\rb,|{\beta \bf K}^{-1}, \rb_c, 0) = \frac{1}{V_{\rm
  box}}.  
\label{eq:p0}
\end{equation}
The probability density of the coupled system at
$\rb=\rb_c$, $P(\rb |{\beta \bf K}^{-1}, \rb_c,1)$ , is simply given
by
\begin{eqnarray}
\lim_{\bf K \rightarrow 0} P(\rb |{\beta \bf K}^{-1}, \rb_c,1)
& = & \frac{e^{-\beta w(\rb_c)}}{
  \int_{V_{\rm box}} e^{-\beta w(\rb) } d\rb }
  \nonumber \\
& = & \frac{1}{{V_T} \left [ 1 + \frac{(V_{\rm box}-V_T)}{V_T} 
e^{\beta w(\rb_c)} \right ]}  = \frac{1}{V_T(1+c_r)} \simeq \frac{1}{V_T} 
\label{eq:k0}
\end{eqnarray}
where the constant $ c_r = \frac{(V_{\rm box}-V_T)}{V_T} e^{\beta
  w(\rb_c)} \simeq \frac{V_{\rm box}}{V_0} e^{-\beta \Delta G_0}$ can
be neglected as long as $e^{\beta \Delta G_0}\gg V_{\rm box}/V_0$.  Plugging
Eqs. \ref{eq:k0} and \ref{eq:p0} into Eq. \ref{eq:COM}, using
Eq. \ref{eq:dg0com} and defining $\Delta G_d(\rm DAM)= \lim_{\bf K
  \rightarrow 0} \Delta G_r({\beta \bf K}^{-1},\rb_c) - \Delta G_u $,
we finally obtain for the unrestrained (DAM) regime
\begin{equation}
  \Delta G_{d0}\simeq \Delta  G_d(\rm DAM)   + k_BT \ln \left ( 
    \frac{V_{\rm box}}{V_0} \right ) 
\label{eq:damdg}
\end{equation}
thus recovering Eq. \ref{eq:fraction} with $\Delta  G_d({\rm DAM}) = 
\Delta G_d(V_{\rm box})$.  
\begin{figure}
\includegraphics[scale=2.0]{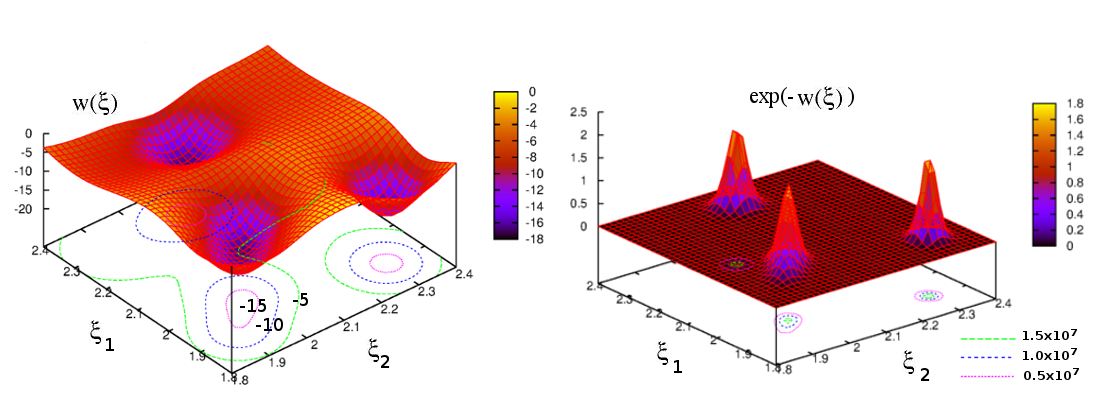}
\caption{Example of a 2D generic PMF with
  multiple minima (left, energy units in $k_BT$) and corresponding
  $e^{-\beta w(\xi)}$ factor using a combination of multivariate
  Gaussian distributions.}
\label{fig:pmf3d}
\end{figure}
I stress that Eq. \ref{eq:damdg} holds only if the MD box volume is
such that $e^{\beta G_0}\gg V_{\rm box}/V_0$.  It should also be
noticed that, while $\Delta G_0$ in Eq. \ref{eq:dg0com} is a purely
{\it conventional} quantity defined with respect to an arbitrarily
selected standard concentration, $\lim_{\bf K \rightarrow 0} \Delta
G_r({\beta \bf K}^{-1},\rb_c)$ and $\Delta G_u$ in Eq. \ref{eq:COM}
refer to free energy differences between two {\it real}
thermodynamic states, namely the decoupling of the {\it unrestrained} ligand in
presence of the receptor in the MD box of volume $V_{\rm box}$ and the
decoupling of the $V_{\rm box}$-independent ligand in the bulk phase,
respectively. If on the rhs of Eq {\ref{eq:k0} we let $V_{\rm box}
  \rightarrow \infty$ , we obtain
\begin{eqnarray}
\lim_{V_{\rm box}\rightarrow \infty}  \frac{1}{{V_T} \left [ 1 + \frac{(V_{\rm box}-V_T)}{V_T} 
e^{\beta w(\rb_c)} \right ]} & = & \frac{e^{-\beta w(\rb_c)}}{V_{\rm box}}
\label{eq:k0lim}
\end{eqnarray}     
Inserting this result and Eq. \ref{eq:p0} into Eq. \ref{eq:COM}, 
an using the definition $\Delta G(\rm DAM) = \lim_{\bf K \rightarrow 0}
\Delta G_r({\beta \bf K}^{-1},\rb_c) - \Delta G_u$,  
we trivially obtain 
\begin{equation}
\lim_{V_{\rm box}\rightarrow \infty}  \Delta G(\rm DAM) = -w(\rb_c) + w(\rb_c) = 0 
\end{equation}
i.e the $V_{\rm box}$-dependent dissociation DAM free energy goes to
zero for $V_{\rm box}\rightarrow \infty$ , or, equivalently the
decoupling free energy of the complex coincides with the decoupling
free energy of the dissociated state. This happens since in the left
branch of the cycle of Figure \ref{fig:cycle}, when the box becomes
exceedingly large ( so that $e^{\beta G_0}\ll V_{\rm box}/V_0$ ) and
provided that the {\it unrestrained} (DAM) transformation of the
complex is ideally done at equilibrium, then ligand in the fully
coupled state at $\lambda=1$ should be found freely wandering in the
bulk with unitary probability, as first remarked in
ref. \cite{karplus03}.  

I conclude this section with some remarks on the nature of $V_T$
appearing in Eqs \ref{eq:dg0com} and \ref{eq:k0}. This quantity has
the unit of a volume and can be identified with the overall
(translational) binding site volume of the ligand ``pose'' on the
protein surface. In order to estimate $V_T$ in a unrestrained
simulation of the complex, one must define, in each sampled bound
configuration, a protein reference frame with respect to which the polar angles
$\theta,\phi$ are evaluated. $V_T$ is hence modulated by the
ro-vibrational coordinates of {\it both} ligand and receptor. For
fluxional ligands and receptors with conformational configurations
widening the COM probability density in the bound state, the pose in
the $\rb$ domain can hence be very rugged indeed as schematically
shown in Figure \ref{fig:pmf3d}.  This picture of the translational
PMF $w(R,\theta,\phi)$ with many crowded competing minima
characterizing the ``pose'' is consistent with the ``induced fit'' or
conformational proofreading model for binding whereby the ligand
and/or the receptor kinetically adjust their conformational states due
to their mutual interaction.\cite{Savir2007}

\subsection{Dissociation free energy via non equilibrium alchemical
  transformation} In spite of the previously outlined wandering ligand
problem, the DAM theory has been used for many years before the advent
of DDM theory, incorporated in popular MD packages\cite{Kollman1995}
and often producing reliable free energy values.\cite{Kollman1993}
Even quite recently,\cite{Yamashita09} Fujitani and coworkers used the
unrestrained DAM and FEP to compute the binding free energy of the
FKBP12-FK506 drug-receptor system.  In all these early DAM
simulations, as well as in the recent examples due to Fujitani and
co-workers,\cite{Yamashita2014,Yamashita2015} the decoupling process
in the left branch of the cycle in Figure \ref{fig:cycle}(a), was
performed, {\it starting from a bound state}, in a total simulation
time (along the whole alchemical decoupling path) never exceeding, at
most, the few tens of nanoseconds. For states with $\lambda$
approaching to zero, the unrestrained ligand could hence easily leave
the binding site and start to freely drift off in the MD box. The time
scale of a random encounter in typical MD box of volume $V_{\rm box}$
containing a single drug-receptor pair can be straightforwardly
estimated from the mean free path, $\frac{V_{\rm box}}{\pi d^2}$ (with
$d$ being the mean radius of the receptor assumed to be much larger
than that of the ligand), and the diffusion coefficient of a ligand in
water,\cite{Wilke1955} typically obtaining collision rates of the order of
0.1:0.01 ns$^{-1}$, i.e. a {\it random} collision every 10 to 100 ns.
In the light of this estimate, we can safely say that {\it all} of the
DAM/FEP or DAM/TI simulations appeared on the literature were actually
non equilibrium processes hence providing {\it a non equilibrium
  estimate} of the decoupling free energy. The same argument applies
to DDM simulations as well, where Boltzmann sampling is in principle
required for all conformational states of the complex that are not
subject to restraints.  Conformational transitions in flexible protein
side chains occur in a wide range of time scale, from picoseconds to
milliseconds and longer.\cite{Miao2016} A converged sampling of these
CVs, for {\it all $\lambda$ states}, that should be highly relevant in
induced fit ligand-receptor association, is in many cases out of the
reach in DDM/FEP or DDM/TI simulations lasting at most few ns per
alchemical state.

In the following, I shall discuss how alchemical non equilibrium
decoupling processes can be used to derive reliable estimates of the
standard dissociation free energies. 
The Jarzynski theorem\cite{jarzynski97} represents one of the few exact
results in non equilibrium thermodynamics, relating the work done in a non
equilibrium (NE) transformation between two thermodynamic states A, B
to the corresponding free energy difference, that is to the work
done {\it reversibly}:
\begin{equation}
e^{\beta \Delta G_{AB}} = \langle e^{-\beta W_{AB}} \rangle_{A}
\label{eq:jt}
\end{equation}
While the configurations of the {\it starting} state A are canonically
sampled, the {\it arrival} configurations of B are not distributed canonically.
The mean NE work, when averaged over many
realizations, all done according to a common prescribed time schedule,
is always larger than the free energy, i.e. the minimum, reversible
work connecting two states. The difference between the average NE work
and the free energy correspond to the mean {\it dissipation} of the NE
process, a function of the speed of the NE realizations.   For
infinitely slow (quasi-static) realizations, the work is always equal
to $\Delta G_{AB}$ and the Jarzynski work average is equivalent to TI,
while for instantaneous processes, it can be shown that that Jarzynski
theorem becomes equivalent to the Zwanzig free energy perturbation
formula. The work probability distributions for the forward (A to B) and reverse
process (B to A) obey the  the Crooks
\begin{equation}
\frac{P_{\rm A\rightarrow B }(W)}{P_{\rm B\leftarrow A}(-W)}  = e^{-\beta (W_{AB}-\Delta F)}
\label{eq:cft}
\end{equation}
The sign of the work in the reverse distribution is due to the fact
that the reverse process is assumed to be done with identical but
inverted time schedule.  It has been
observed\cite{Goette2009,Procacci14,Gapsys12} that the work
distribution obtained from fast annihilation/creation NE processes
(lasting no more than few hundreds or even tens of picoseconds) of
small to moderate size organic molecules in polar non polar solvents
has a marked Gaussian character and that the corresponding dissipation
is surprisingly small, ranging from 0.05to 0.1 kcal mol$^{-1}$ per atom.
In case of Gaussian work distributions for the (forward) annihilation
process, the Crooks theorem, Eq. \ref{eq:cft}, provides an {\it
  unbiased} estimate of the free energy in the form of
\begin{equation}
\Delta G = \langle W_{A \rightarrow B}  \rangle - \frac{\beta \sigma^2}{2}  
\label{eq:jarzy2}
\end{equation}
where $\langle W_{A \rightarrow B} \rangle$ and $\sigma$ are the mean
work and variance of many NE realizations. This fact has been recently
exploited\cite{fsdam,pccp1} to implement a non equilibrium approach to
alchemical simulation.  In this methodology the dissociation free
energy is again accessed via the thermodynamic cycle, but this time
the annihilation processes on the two branches are done irreversibly
at fast speed, starting form the fully coupled equilibrated
states. The free energy is recovered either from the Jarzynski
theorem, Eq. \ref{eq:jt} or, in case of Gaussian work distribution,
from the unbiased estimate, Eq. \ref{eq:jarzy2}. As such, the NE
alchemical variant is compatible either with the version with strong
or weak restraints or with the unrestrained approach.  In case the NE
alchemical simulations with restraints, the quantities $\Delta
G_r(\SB_{r},\rb_c,\xib_c)$ and in $ \Delta G_u(\xib_c)$ in
Eq. \ref{eq:DGD} or $\Delta G_r(\SB_{r},\rb_c,\OB_c)$ and in $ \Delta
G_u(\OB_c)$ in Eq. \ref{eq:gilson} are not evaluated using TI or FEP;
rather they are computed applying Eq. \ref{eq:jarzy2} or Eq. \ref{eq:jt}
to the work histograms obtained by launching in parallel few hundreds
of fast (0.1 to 0.5 ns) decoupling alchemical independent
trajectories. For the unrestrained (DAM) NE version, 
it has been shown\cite{pccp1,pccp2} that the dissociation free energy
can be recovered exploiting the Crooks theorem applied to mixture of
Gaussian distributions, landing on Eq.
\begin{equation}
\Delta G_0 = \Delta G_{b} - \Delta G_{\rm u} +  k_BT \ln
\frac{V_{site}}{V_0}
\label{eq:dam}
\end{equation}
where $\Delta G_{b}$ is NE free energy Gaussian estimate for the fast
annihilation of the bound state, $\Delta G_{u}$ is NE free energy
estimate for the fast annihilation of the ligand in and $V_{\rm
  site}$ should correspond to the effective cumulative ``volume'' of
the binding site or, using a definition due again to
Gilson\cite{Gilson2004}, to the {\it exclusion zone} of the receptor,
defined by a measurable (in principle) probability of re-entrance in
an hypothetical reverse process for the complex.  As long as the NE
process is much faster compared to the time scale of the relative
ligand-receptor diffusion, the NE estimate $\Delta G_{b}$ via
Eq. \ref{eq:jt} or \ref{eq:jarzy2} is essentially independent of the
box volume and on the duration time of the NE process, so that we can
identify $V_{\rm site}$ with $V_{T}$ in Eq. \ref{eq:damdg}. This is a
rather trivial consequence of the insensitivity of the equilibrium
constant integral $K_e=\int_{{\cal D}_b} e^{-\beta w(\rb}) d\rb $ to
the integration domain ${\cal D}_b$ defining the region of existence
of the complex and to the fact that in the fast switching alchemical
decoupling of the bound state, the decoupled ligand does not have the
time to explore regions that are far away from ${\cal D}_b$.

The NE alchemical approach, whether in the restrained or unrestrained
version, bypass completely the need for an equilibrium sampling at the
intermediate alchemical states, requiring a canonical sampling only at
starting fully coupled $\lambda=1$ thermodynamic state. The latter can
be obtained using enhanced sampling techniques such as H-REM or
Umbrella Sampling.\cite{edu} With this regard, the apparent ability of
{\it equilibrium} FEP or TI based approaches to produce reliable
estimates of the binding free energy in conventional simulation
lasting few ns per alchemical states (i.e. for a timescale that is
well below the characteristic ergodicity timescale in drug-receptor
systems) is actually a fortuitous consequence of non equilibrium
processes.  These techniques are in fact unaware applications of non
equilibrium approaches whereby a mean alchemical {\it work}, rather
than a free energy, is determined. Such work, if the alchemical
process is done in a cumulative time of the order of the tens of
nanoseconds, is Gaussianly distributed over few $k_BT$ or less and, in
force of the Crooks theorem, must be close to the true decoupling free
energy. The similarity of the dissipation energy on the two branch of
the cycles provides a further fortuitous compensation effect when
evaluating the dissociation free energy as a difference of two non
equilibrium mean work.

\section{Conclusions}
In this paper I have revisited the statistical mechanics of non
covalent bonding in drug-receptor systems. I have shown that all
existing alchemical theories in binding free energy calculations can
be rationalized in term of a unifying treatment encompassing the
original unrestrained DAM\cite{jorgensen85}, the Gilson's restrained
DDM variant\cite{gilson97} and the sophisticated docking approach
proposed by Deng and Roux.\cite{Deng2006} The cited alchemical
theories differ in the definition (explicit or implicit) of the binding
site volume through the enforcement of a set of appropriately selected
restrained potentials.
Strong restrained approaches\cite{Deng2006} relies on a
precise knowledge of the binding pose and volume in the context of the
traditional picture of the lock and key model. The DDM and DAM
theories make weaker assumptions on the pose topology and nature,
hence being progressively shifted towards a more realistic induced
fit/ conformational proofreading model in drug-receptor interaction.
All alchemical theories are finally placed into the broader context
of non equilibrium thermodynamics, discussing the application of the
Crooks and Jarzynski non equilibrium theorems to the evaluation of
alchemical decoupling free energies.

\bibliography{ms}
\end{document}